\documentclass[aps,pre,amsmath,amssymb,reprint,numeric,fleqn]{revtex4-1}

\usepackage{comment}
\usepackage{braket}
\usepackage{xcolor}

\usepackage[english]{babel}
\usepackage{float}
\usepackage{dcolumn}
\usepackage[applemac]{inputenc}
\usepackage{color}
\usepackage{bm}
\usepackage[normalem]{ulem}
\usepackage{multirow}
\usepackage{lipsum}
\usepackage{graphicx}
\usepackage{grffile}
\usepackage{subfigure}

\begin{document}

\title{Diffusion Properties of a Brownian Ratchet with Coulomb~Friction}

\author{Massimiliano Semeraro$^{1}$, Giuseppe Gonnella$^{1}$, Eugenio Lippiello$^{2}$ and Alessandro Sarracino$^{3,}$}

\affiliation{$^{1}$Dipartimento Interateneo di  Fisica,  Universit\`a  degli  Studi  di  Bari  and  INFN,
Sezione  di  Bari,  via  Amendola  173, I-70126~Bari,  Italy;\\
$^{2}$Dipartimento di Matematica e Fisica, Universit\`a della Campania ``Luigi Vanvitelli'', 81100~Caserta, Italy;\\
$^{3}$Dipartimento di Ingegneria, Universit\`a della Campania ``Luigi Vanvitelli'', 81031~Aversa, Italy
}

\begin{abstract}
  The motion of a Brownian particle in the presence of Coulomb
  friction and an asymmetric spatial potential was evaluated in this
  study. The system exhibits a ratchet effect, i.e., an average
  directed motion even in the absence of an external force, induced by
  the coupling of non-equilibrium conditions with the spatial
  asymmetry. Both the average motion and the fluctuations of the
  Brownian particle were analysed. The stationary velocity shows a
  non-monotonic behaviour as a function of both the temperature and
  the viscosity of the bath.  The diffusion properties of the
  particle, which show several time regimes, were also
  investigated. To highlight the role of non-linear friction in the
  dynamics, a comparison is presented with a linear model of a
  Brownian particle driven by a constant external force, which allows
  for analytical treatment. In particular, the study unveils that the
  passage times between different temporal regimes are strongly
  affected by the presence of Coulomb friction.
\end{abstract}

\maketitle

\section{Introduction}

Ratchet models (or Brownian motors) are systems where non-equilibrium conditions can be exploited to extract work from random fluctuations~\cite{hanggi2009artificial}. Due to the breaking of temporal and spatial symmetries, even in the absence of an external drive, these systems present a spontaneous average net drift, which would be forbidden in equilibrium conditions. Several different sources of non-equilibrium dynamics can be considered: time-dependent forcing, as in flashing ratchet~\cite{reimann2002introduction}; correlated noise~\cite{dialynas1997ratchet}; slow relaxation in glasses~\cite{gradenigo2010ratchet}; dissipative interactions as in granular systems~\cite{cleuren2008dynamical,costantini2007granular}; the presence of velocity-dependent forces~\cite{sarracino2013time}; and even the self propulsion in active matter systems~\cite{arjun2022}. 

An intriguing example of a force that depends on the velocity of the particle is represented by Coulomb (or dry) friction, which takes into account the energy dissipation contribution due to the slipping on a surface. This force can be introduced into a Langevin equation as a constant-magnitude force whose sign is opposite to the particle velocity.
The interest in this model was first raised by de Gennes in one of his
late papers~\cite{de2005brownian} and by Hayakawa in~\cite{hayakawa2005langevin}. As mentioned in~\cite{de2005brownian}, examples of physically relevant situations where the interplay between Coulomb friction and Brownian motion can have an interesting role are a micron-size solid particle under thermal noise and a macroscopic particle on a vibrated surface. The stochastic equation for the particle velocity under the action of dry friction has been widely studied, and some analytical results have been also obtained
in the absence of spatial potential, in particular, via a path integral approach~\cite{baule2009path},  from the Fokker--Planck equation that can be solved to obtain the time-dependent propagator and the particle velocity correlation function~\cite{touchette2010brownian}, or even in the presence of an external force~\cite{baule2010stick}, or in periodically driven systems~\cite{pototsky2013periodically} and in the presence of an elastic band~\cite{feghhi2022pulling}
. Other studies have focused on the issues related to the definition of entropy production in these systems~\cite{cerino2015entropy}. In the specific context of models of Brownian motors, the role of Coulomb friction as a source of non-equilibrium able to induce a ratchet effect has  also been investigated in different systems ~\cite{sarracino2013time,sarracino2013ratchet,manacorda2014coulomb}, with experimental realizations in the context of driven granular gases~\cite{gnoli2013brownian,gnoli2013granular}. 

As mentioned above, in order to induce a directed motion, an asymmetric spatial potential is crucial. This introduces a coupling between positions and velocities, making the problem not analytically tractable. 
Here, {this case is studied with extensive numerical simulations with a focus on the dynamics of an underdamped Langevin equation in the presence of an asymmetric periodic potential and Coulomb friction}. In particular, in Section~\ref{velocity}, { the model and its main parameters are introduced. The average ratchet velocity is investigated} as a function of the viscosity and of the temperature of the thermal bath, showing that there are optimal values maximising the ratchet effect. In Section~\ref{diff}, { the diffusion properties of the system are considered}, investigating the behaviour of the position variance and the mean square displacement (MSD) for a wide range of parameters. { A simple diffusive behaviour at large times is found in the variance}, while a more complex scenario is observed for the MSD due to the presence of different time regimes. In Section~\ref{constantforce}, { a comparison is presented of some of the observed trends}, with those relative to an analytically solvable model consisting of an underdamped Brownian particle driven by a constant external force. In order to further deepen the system behaviour,
{ the effect of the Coulomb friction on the average first exit time from a parabolic potential well is also investigated}. Finally, in Section~\ref{concl}, { a summary of  and comments on our findings are presented}.  

\section{Langevin Equation with Coulomb Friction and Ratchet Effect}
\label{velocity}

{ The system consists of} a unitary mass inertial particle in one dimension in contact with a thermal bath in the presence of both an asymmetric spatial potential and a nonlinear velocity-dependent friction force. The model is described by the following Langevin equation
\begin{equation}\label{model}
\begin{cases}
&\dot{x}(t)=v(t)\\
&\dot{v}(t)=-\gamma v(t)-U'[x(t)]-\alpha \sigma[v(t)]+\sqrt{2\gamma T}~\xi(t),
\end{cases}
\end{equation}
where $x(t)$ and $v(t)$ are the position and velocity of the particle, respectively; $\gamma$ is the viscous friction coefficient; $U[x(t)]$ is an external potential (the prime denoting a derivative with respect to $x$); $\alpha$ is the constant amplitude of the Coulomb friction; $\sigma(v)$ is the sign function ($\sigma(0)=0$); $\xi(t)$ is white noise with $\braket{\xi(t)}=0$ and $\braket{\xi(t)\xi(t')}=\delta(t-t')$; and $T$ is the bath temperature {(we take the Boltzmann constant $k_B=1$ throughout the manuscript)}. The spatial potential is chosen as the ratchet potential
\begin{equation}\label{potential}
    U[x(t)]=\sin[x(t)]+\mu\sin[2x(t)],
\end{equation}
where $\mu$ is the parameter that introduces the spatial asymmetry. { Note} that the term $\sin(2x)$ keeps the potential periodic. For $\mu=0$ the potential is symmetric, and no ratchet effect occurs. This model was first studied in~\cite{sarracino2013time}, where different velocity-dependent friction forces were considered and the form of the non-equilibrium generalized fluctuation-dissipation relation was investigated. 

As shown in~\cite{sarracino2013time}, the system \eqref{model} does not satisfy the detailed balance condition due to the presence of both the nonlinear velocity-dependent dry friction and of the ratchet potential \eqref{potential}.
As discussed in~\cite{dubkov2009non}, the issue of recovering detailed balance in Langevin equations with non-linear velocity-dependent forces requires the introduction of a non-Gaussian thermal bath and a multiplicative noise. {The out-of-equilibrium dynamics of the system can instead be  exploited} to induce the ratchet effect, namely a finite non-zero particle average velocity $\braket{v(t)}$ in the stationary state.

In the following, {the analysis obtained from the numerical integration of the stochastic differential Equation \eqref{model} via the Euler--Maruyama algorithm~\cite{platen2010} with time step $dt=10^{-3}$ is presented. The code has been written in the Python programming language. The stability of results for $dt \leq 10^{-3}$ has been checked.} The simulation's reduced units are provided {as follows: the potential periodicity $\Delta x=2\pi$ is the space unit, the potential depth $\Delta U$ is the energy unit, and the inverse of the friction coefficient $\gamma^{-1}$ is the time unit; the velocity unit is given by $\gamma \Delta x$.} In \cite{sarracino2013time}, the stationary velocity of the particle was studied at fixed values of the {friction coefficient $\gamma=0.05$ and temperature $T=10$}, finding a maximum for  $\mu=0.4$ and $\alpha = 1$. Here, we extend the investigation of the model and present the complementary analysis by fixing the asymmetry parameter and the amplitude of dry friction and exploring a wide spectrum of $T$ and $\gamma$ values.

\begin{figure}[H]
\includegraphics[width=1.0\linewidth]{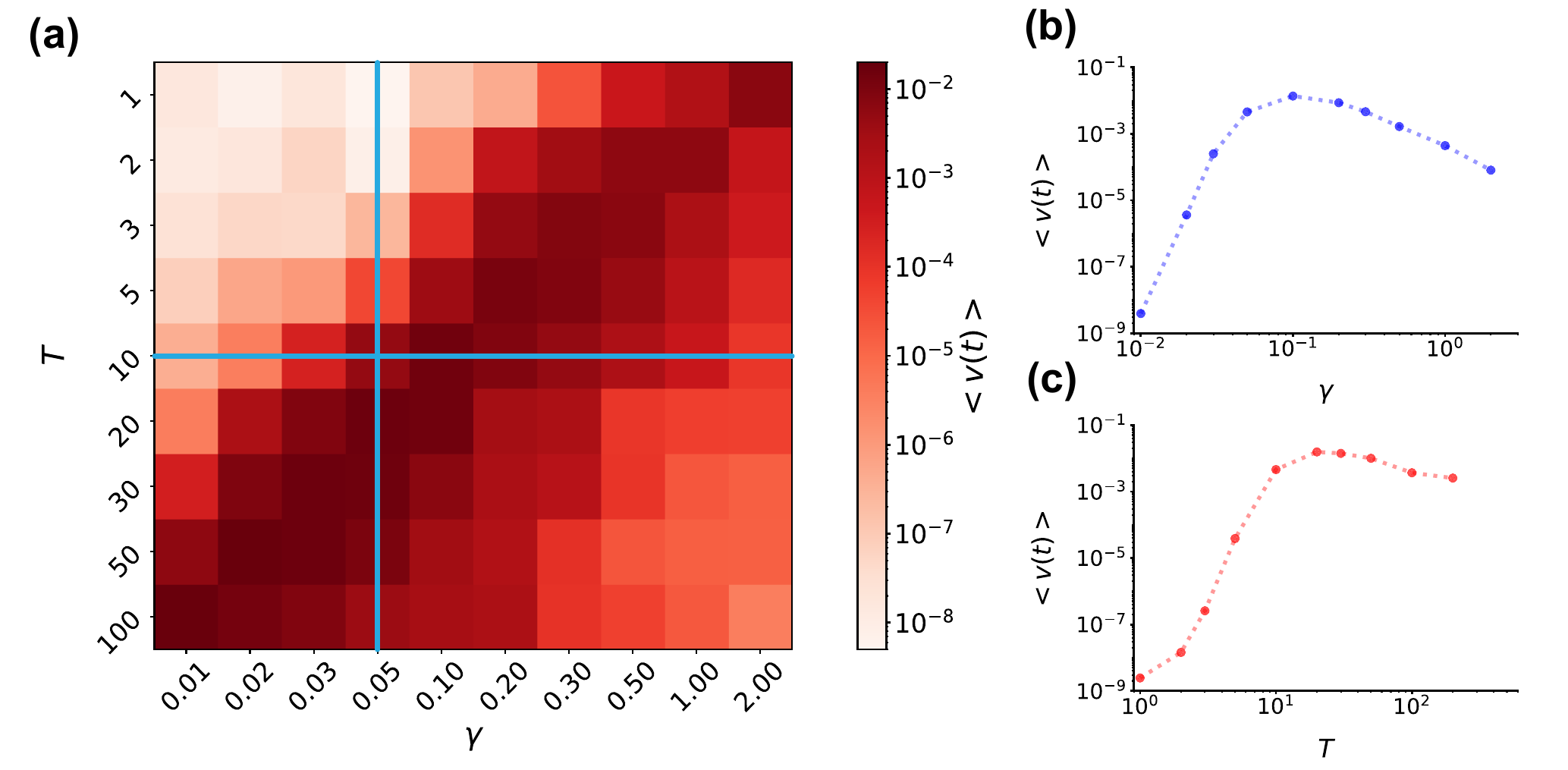}
\vspace{-0.25cm}
\caption{
  \footnotesize{{\bf Average ratchet drift
  vs.  temperature $T$ and friction coefficient $\gamma$.}
  (\textbf{a}): Contour plot for the average ratchet velocity
  $\braket{v(t)}$ as a function of the temperature $T$ and the
  friction $\gamma$. Parameters: $\alpha=1.0$, $\mu=0.4$. The blue
  horizontal and vertical lines highlight the $\braket{v(t)}$ values
  at fixed $T$ and $\gamma$ used to plot panels (\textbf{b}) and
  (\textbf{c}), respectively. (\textbf{b}) and (\textbf{c}): Trend of
  $\braket{v(t)}$ as a function of $\gamma$ at fixed $T=10.0$ and $T$
  at fixed $\gamma=0.05$, respectively, as extracted from panel
  (\textbf{a}).}}
\label{fig:fig1}
\end{figure}

Results are presented in the contour plot of Figure~\ref{fig:fig1}a. Quite interestingly, {a non-monotonic behaviour of movement along both the temperature $T$ and the friction coefficient $\gamma$ axes (see columns and rows) is evident}. This observation is further clarified by panels (b) and (c), reporting the average velocity as a function of $\gamma$ and $T$ at fixed $T$ and $\gamma$, respectively, extracted from the row and column of the chart in panel (a) denoted by the blue lines. This means that optimal choices of these parameters, such that the system reaches its maximal velocity and the ratchet effect is most enhanced, are possible. In particular, the figure reveals that at fixed $\mu$ and $\alpha$, any choice such that $\gamma T \sim 1$  (see darker diagonal) results in a maximum ratchet effect. The explanation behind this non-trivial phenomenon relies on the two-fold role played by the temperature: on the one hand, thermal fluctuations allow the particle to explore the spatial potential so that the ratchet effect can actually take place; on the other had, for too-large values of $T$, the white noise is enhanced and the particle dynamics is less affected by the presence of the spatial potential, thereby damping the ratchet mechanism. {The friction coefficient $\gamma$ plays a similar two-fold role: it contributes to the amplitude of the random force, as does the temperature, allowing the particle to explore the periodic potential, but it also represents the viscous friction experienced by the particle,
which hinders the dynamics.}

\section{Diffusion Properties}
\label{diff}

Here, {the analysis of the fluctuations around the average motion and of the diffusion properties is presented. These quantities are relevant} to highlighting the role of dry friction in the dynamics of the Brownian particle. We focus on the position variance $Var[x(t)]=\langle x(t)^2\rangle-\langle x(t)\rangle^2$ and the MSD $\Delta(t,t_0)=\langle[x(t)-x(t_0)]^2\rangle$ .

\begin{figure}[H]
\includegraphics[width=1.0\linewidth]{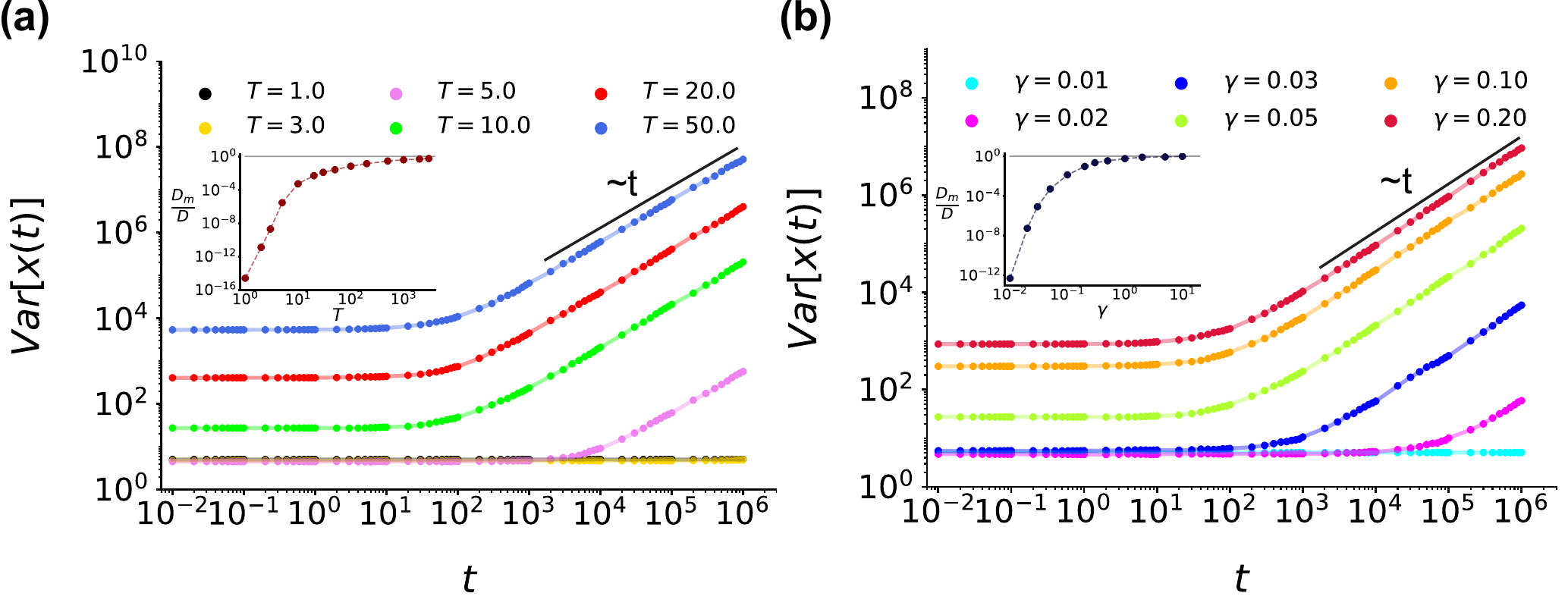}
\vspace{-0.25cm}
\caption{
\footnotesize{{\bf Variance 
 vs. {temperature $T$ and friction coefficient} $\gamma$.} Position variance $Var[x(t)]$ for various choices of temperature $T$ at fixed $\gamma=0.05$ (panel (\textbf{a})) and for various choices of friction coefficient $\gamma$ at fixed $T=10$ (panel (\textbf{b})) up to simulation time $t=10^6$. The insets report the ratio between the measured diffusion coefficient $D_m$ and the asymptotic underdamped one $D=T/\gamma$. Other parameters: $\alpha=1.0$, $\mu=0.4$.}}
\label{fig:fig2_T_gamma}
\end{figure}

The position variance $Var[x(t)]$ is reported in Figure~\ref{fig:fig2_T_gamma}, for various choices of {temperature $T$ (panel (a)) and friction coefficient $\gamma$} (panel (b)) at fixed $\gamma$ and $T$, respectively. The variance is stuck to a constant parameter-dependent value at small times as the particle is trapped in the potential well. This phenomenon resembles the dispersionless diffusion regime described in~\cite{lindenberg2007dispersionless}, which has recently been reconsidered in~\cite{kaffashnia2021origin,marchenko2022dispersionless}. If the system parameters allow the particle to eventually escape from the initial potential well, it starts exploring other regions, and a diffusive regime $Var[x(t)]\sim t$ sets in. The diffusion coefficient $D_m$ characterising such a regime was measured and is reported in the insets of Figure~\ref{fig:fig2_T_gamma}. As intuitively expected, $D_m$ shows a monotonic trend with $T$ and $\gamma$, from low values due to the trapping action of the potential  towards the overdamped prediction $D=T/\gamma$ in both cases (see insets of panels (a) and (b)).

\begin{figure}[H]
\includegraphics[width=1.0\linewidth]{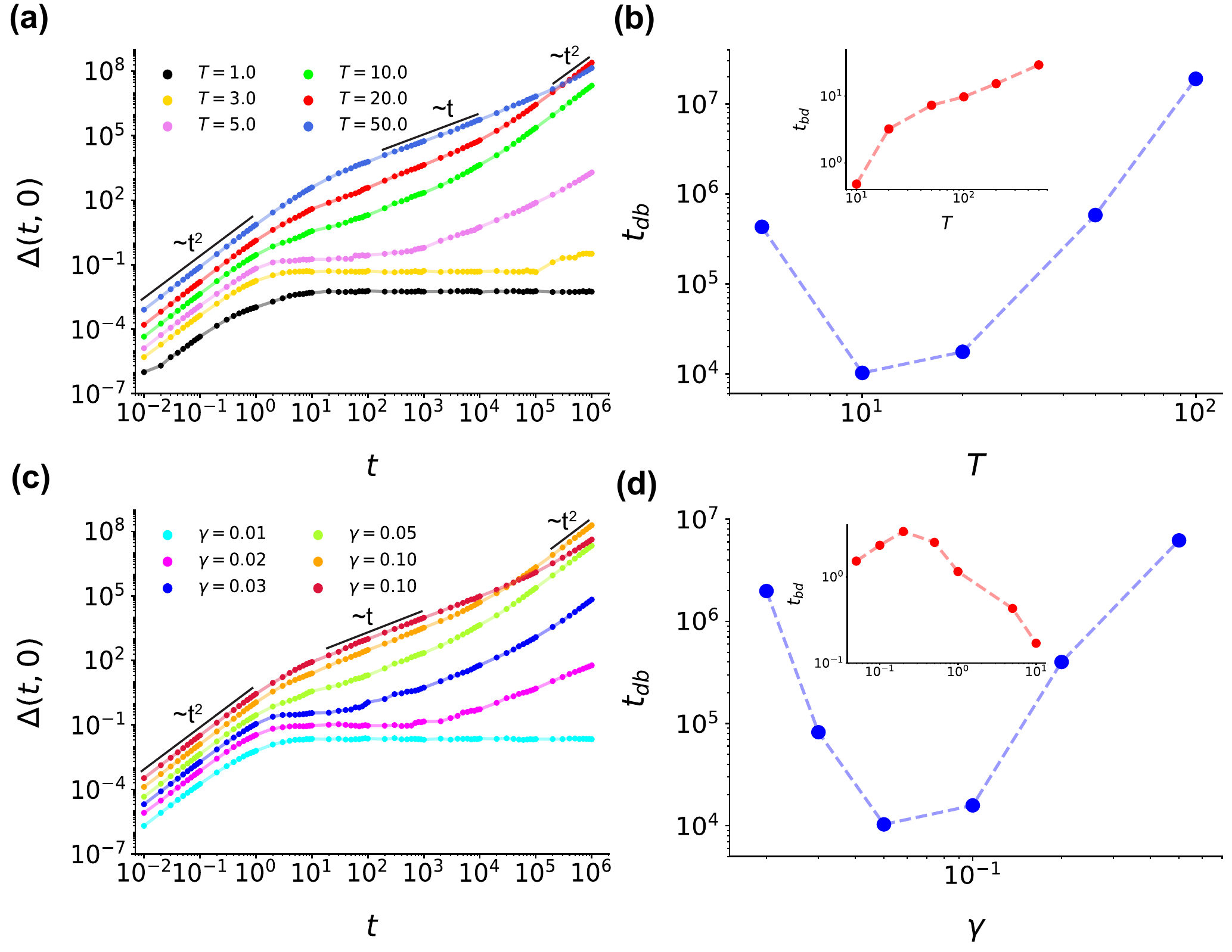}
\vspace{-0.25cm}
\caption{
\footnotesize{{\bf Diffusion
 properties vs. {temperature $T$ and friction coefficient} $\gamma$.} (\textbf{a}) and (\textbf{c}): Position mean square displacement $\Delta(t,0)$ for various choices of temperature $T$ at fixed $\gamma=0.05$ (panel (a)) and friction coefficient $\gamma$ at fixed $T=10.0$ (panel (c)) up to simulation time $t=10^6$. (\textbf{b}) and (\textbf{d}): Diffusive$\rightarrow$ballistic crossover times $t_{db}$ as a function of the temperature $T$ at fixed $\gamma=0.05$ (panel (b)) and friction coefficient $\gamma$  at fixed $T=10.0$ (panel (d)) measured from $\Delta(t,0)$. The insets report the measured ballistic$\rightarrow$diffusive crossover times $t_{bd}$. Other parameters: $\alpha=1.0$, $\mu=0.4$.}}
\label{fig:fig3_T gamma}
\end{figure}

The MSD  instead shows a plateau, followed by diffusive and ballistic regimes due to the interplay between particle fluctuation and ratchet potential. As shown in Figure~\ref{fig:fig3_T gamma}a,c one can identify several regimes: (i) a first initial ballistic regime due to inertial effects; (ii) a plateau regime, whose duration depends on the values of {friction coefficient $\gamma$ and temperature} $T$; (iii) a diffusive regime; (iv) a final ballistic regime due to the directed motion induced by the ratchet effect. In order to analyse the dependence of the time duration of such regimes on the model parameters, we focus on the crossover times for the passage from ballistic to diffusive $t_{bd}$ and from diffusive to ballistic $t_{db}$ extracted from the intercept between the curves $\sim t$ or $\sim t^2$ fitting two consecutive regimes. The measured values can be appreciated from Figure~\ref{fig:fig3_T gamma}b,d. An interesting non-monotonic behaviour of $t_{db}$ as function of both $T$ and $\gamma$ appears. As already underlined, this occurs because the temperature increase above a certain threshold allows thermal fluctuations to play a dominant role, making  the spatial potential less effective and hindering the occurrence of the ratchet mechanism. This therefore leads to a larger time for the final ballistic regime to set in. Regarding the behaviour of the crossover time from the initial ballistic regime due to inertia to the intermediate diffusive regime, from the inset of Figure~\ref{fig:fig3_T gamma}b, {one observes} a continuous growth as a function of $T$, while from the inset of Figure~\ref{fig:fig3_T gamma}d, a non-monotonic trend as function of $\gamma$ is apparent. Indeed, as the viscous friction $\gamma$ is increased at fixed temperature, the inertial effects become negligible, making the passage to diffusion very rapid. On the contrary, an increase in $T$ at fixed $\gamma$ results in a longer inertial regime. These behaviours will be reconsidered in the light of the simple constant force model discussed in Section~\ref{constantforce}.  Finally, {note that  for the range of parameters investigated, one can clearly identify} the first crossover time $t_{bd}$ only in a few cases, so  {one cannot} comment on its general behaviour.    

\section{Constant Force Model}
\label{constantforce}

The ratchet effect is characterised by a spontaneous net drift arising from the coupling of non-equilibrium conditions with spatial asymmetry. A net average velocity can also be  trivially induced with only viscous friction and no spatial potential, applying a constant external force. Here {this simple case is considered}, which allows for analytical treatment and comparison of its diffusional properties with those observed in the ratchet system
. {One finds} that some qualitative features, such as the several MSD regimes, can be reproduced, while others cannot, because of the peculiar role of nonlinear velocity-dependent forces. 

{The constant force model consists of} the following underdamped Langevin equation:
\begin{equation}\label{cfm}
\begin{cases}
&\dot{x}(t)=v(t)\\
&\dot{v}(t)=-\gamma v(t)+F+\sqrt{2\gamma T}~\xi(t),
\end{cases}
\end{equation}
where $F$ is a constant external force.
This can be easily solved, yielding, for the mean velocity and position,
\begin{eqnarray}
      	\braket{v(t)}&=&
    	v_0e^{-\gamma t}+\int_0^t dt' e^{-\gamma(t-t')}(F+\sqrt{2\gamma T}\braket{\xi(t')})\nonumber \\
    	&=&v_0e^{-\gamma t}+Fe^{-\gamma t}\int_0^t dt' e^{\gamma t'} \nonumber\\
    	&=&
    	v_0e^{-\gamma t}+\frac{F}{\gamma}(1-e^{-\gamma t})\rightarrow \frac{F}{\gamma}~,
 	\label{eq:meanvel_Fcost}
\end{eqnarray}
and
\begin{eqnarray*}
    \langle x(t)\rangle &=&
    \Braket{\int_0^t dt'~v(t')}=\int_0^tdt'~\braket{v(t')}\nonumber \\
    &=&\int_0^t dt'~\left[\left(v_0-\frac{F}{\gamma}\right)e^{-\gamma t'}+\frac{F}{\gamma}\right]\nonumber\\
                 &=&
                 \left(v_0-\frac{F}{\gamma}\right)\frac{1-e^{-\gamma t}}{\gamma}+\frac{F}{\gamma}t\rightarrow \frac{1}{\gamma}\left(v_0-\frac{F}{\gamma}\right)+\frac{F}{\gamma}t~,
\end{eqnarray*}
respectively, where the arrows denote the large time limit. In order to compare such behaviors with those found in the ratchet model, for each choice of { temperature $T$ and friction coefficient} $\gamma$, {the constant force is set} at $F=\gamma\braket{v(t)}$, where $\braket{v(t)}$ is the average velocity in the corresponding ratchet system. The velocity auto-correlation function is given by
{
\begin{eqnarray*}
     \braket{v(t_1)v(t_2)}&=&
     v_0^2e^{-\gamma(t_1+t_2)}+\frac{v_0F}{\gamma}{e^{-\gamma t_1}}(1-e^{-\gamma t_2})\nonumber \\
     &+&\frac{v_0F}{\gamma}{e^{-\gamma t_2}}(1-e^{-\gamma t_1})\nonumber \\
     &+&\frac{F^2}{\gamma^2}(1-e^{-\gamma t_1})(1-e^{-\gamma t_2}) \nonumber \\
     &+&T(e^{-\gamma|t_1-t_2|}-e^{-\gamma(t_1+t_2)}),
\end{eqnarray*}
}
so that, at equal times $t_1=t_2=t$, one obtains
\begin{equation*}
        \braket{v^2(t)}\longrightarrow \frac{F^2}{\gamma^2}+T.
\end{equation*}
Finally, the time integration of the velocity autocorrelation function yields the MSD
\vspace{-8pt} 
{
\begin{eqnarray}
        &&\braket{[x(t)-x(0)]^2}=\Braket{\int_0^t v(t')dt'\int_0^2 v(t'')dt''}\nonumber \\
        &=&\frac{v_0^2}{\gamma^2}(e^{-\gamma t}-1)^2+\frac{2v_0F}{\gamma}\left[\frac{1-e^{-\gamma t}}{\gamma}t-\frac{(e^{-\gamma t-}-1)^2}{\gamma^2}\right]\nonumber \\
  &+&\frac{F^2}{\gamma^2}\left[t^2+2\frac{e^{-\gamma t}-1}{\gamma}t+\frac{(e^{-\gamma t}-1)^2}{\gamma^2}\right] \nonumber \\
  &+&\frac{2T}{\gamma}\left(t-\frac{1-e^{-\gamma t}}{\gamma}\right)-T\frac{(e^{-\gamma t}-1)^2}{\gamma^2}.
\label{eq:msd_F_app}
\end{eqnarray}
}
It is interesting to simplify the above expression in the large and small time limits.
At large times $t\gg \gamma^{-1}$, the MSD can be approximated as
\begin{equation*}
    \begin{split}
         \braket{(x(t)-x(0))^2}&\simeq\frac{v_0^2}{\gamma^2}-\frac{2v_0 F}{\gamma^3}+\frac{F^2}{\gamma^4}-\frac{3T}{\gamma^2}\\
         &+2\left(\frac{v_0F}{\gamma^2}-\frac{F^2}{\gamma^3}+\frac{T}{\gamma}\right)t+\frac{F^2}{\gamma^2}t^2.
    \end{split}
\end{equation*}
In the opposite limit, $t\ll \/\gamma$, the exponential expansion around $t=0$ leads to the expression
\begin{equation*}
     \braket{(x(t)-x(0))^2}\simeq v_0^2 t^2+\left(-\gamma v_0^2+v_0F+\frac{2}{3}\gamma T\right)t^3+o(t^4).
\end{equation*}
The previous formulae allow the crossover times to be estimated explicitly. Indeed, one easily finds that at large times, the passage from a diffusive to the ballistic regime occurs at a time
\begin{equation}
    t_{db}=\frac{\gamma^2}{F^2}\left(\frac{v_0F}{\gamma^2}-\frac{F^2}{\gamma^3}+\frac{T}{\gamma}\right)~.
    \label{eq:tdb_Fcost}
\end{equation}
At small times, when $v_0\neq 0$, one instead finds that the initial ballistic regime changes to diffusion at time
\begin{equation*}
    t_{bd}=\frac{v_0^2}{-\gamma v_0^2+v_0F+\frac{2}{3}T}.
\end{equation*}

\begin{figure}[H]
\includegraphics[width=1.\linewidth]{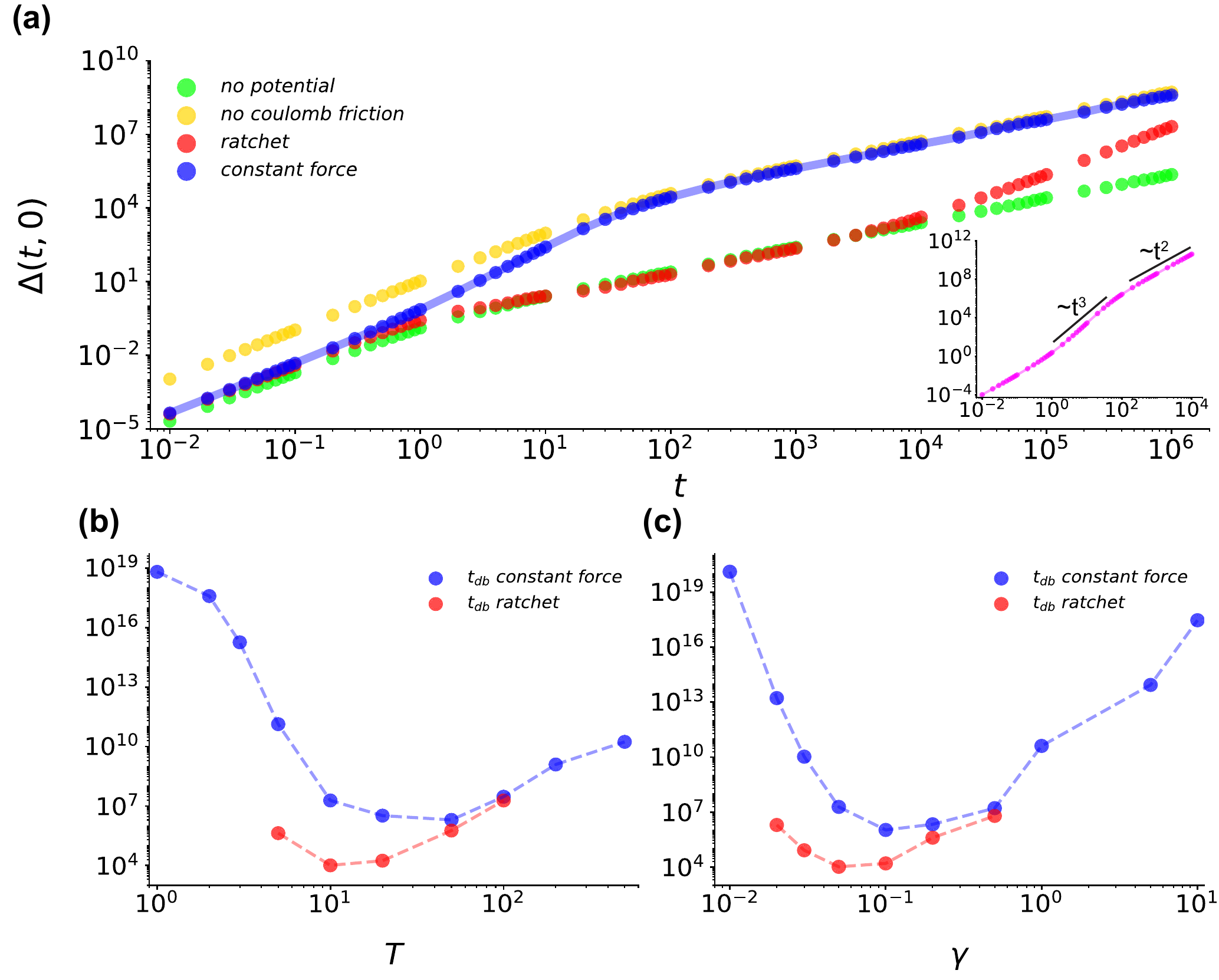}
\vspace{-0.25cm}
\caption{
\footnotesize{{\bf Constant-force model.} (\textbf{a}): Mean square displacement $\Delta(t,0)$ of the constant force model {Equation (\ref{cfm})} compared to those {computed} in the no-potential ({$U=0$ in Equation (\ref{model})), no-Coulomb friction ($\alpha=0$ in Equation (\ref{model})), and ratchet model (Equation (\ref{model}) with $\alpha=1$)}. Points denote numerical results, and the blue solid line is the theoretical expression Equation~\ref{eq:msd_F_app}. The inset reports the theoretical and numerical $\Delta(t,0)$ for a different force, highlighting the $\sim t^3$ superdiffusive  regime. (\textbf{b}) and (\textbf{c}): Comparison between the {computed} $t_{db}$ in the ratchet and in the constant force model as a function of the temperature $T$ and of the friction coefficient $\gamma$, respectively. The blue dots are evaluated through \eqref{eq:tdb_Fcost}, and red dots are numerically estimated. Parameters: $\alpha=1.0$, $\mu=0.4$, $\gamma=10.0$ in panel (b); $T=0.05$ in panel (c). $F=2.28\cdot 10^{-4}$ is the constant force corresponding to the $T=10,  \gamma=0.05, \mu=0.4$ ratchet case, while $F=1.0, v_0=1.0$ are the parameters chosen for the inset.}}
\label{fig:fig4}
\end{figure}

In order to highlight the different contributions from the non-linear dry friction and from the potential, we report in Figure~\ref{fig:fig4}a the behaviour of the MSD for the constant force model compared to the ratchet system. We also compute the MSD in the case $\alpha =0$ (no Coulomb friction) and in the case $U=0$ (no potential), in order to better understand the role of the different terms contributing to the dynamics in the Langevin equation. In the absence of potential, {one observes} that the first two time regimes are very similar to the case of the ratchet model, while, as expected, the final ballistic regime does not take place. On the contrary, in the absence of dry friction, the particle shows a diffusive behaviour similar to the constant force model. Figures \ref{fig:fig4}b,c propose instead a comparison between the trend of the crossover times $t_{db}$ in the constant force model, obtained through \eqref{eq:tdb_Fcost}, and the one in the ratchet model, which  is  numerically evaluated. In all cases, {one finds} a qualitative but not quantitative agreement, with some differences worthy of attention. For example, in some cases, as the ratchet model leaves the initial ballistic regime and enters the diffusive one, the constant force model  instead enters a superdiffusive regime $\sim t^3$ lasting roughly two decades before reaching the final ballistic regime (see inset of panel (a)). An interesting observation regards the onset of the ballistic behaviour as a function of both $T$ and $\gamma$: from Figures \ref{fig:fig4}b,c, one finds that, for small values of { temperature $T$ and viscous coefficient} $\gamma$, respectively, the crossover times from diffusive to ballistic regimes are an order of magnitude smaller in the ratchet model with respect to the constant force model. This reveals the dramatic role played by non-linear friction in speeding the dynamics of the system for a range of parameters. 

\subsection*{Characteristic Escape Times from a Single Well}
\label{exit_times}

\begin{figure}[H]
\includegraphics[width=1.\linewidth]{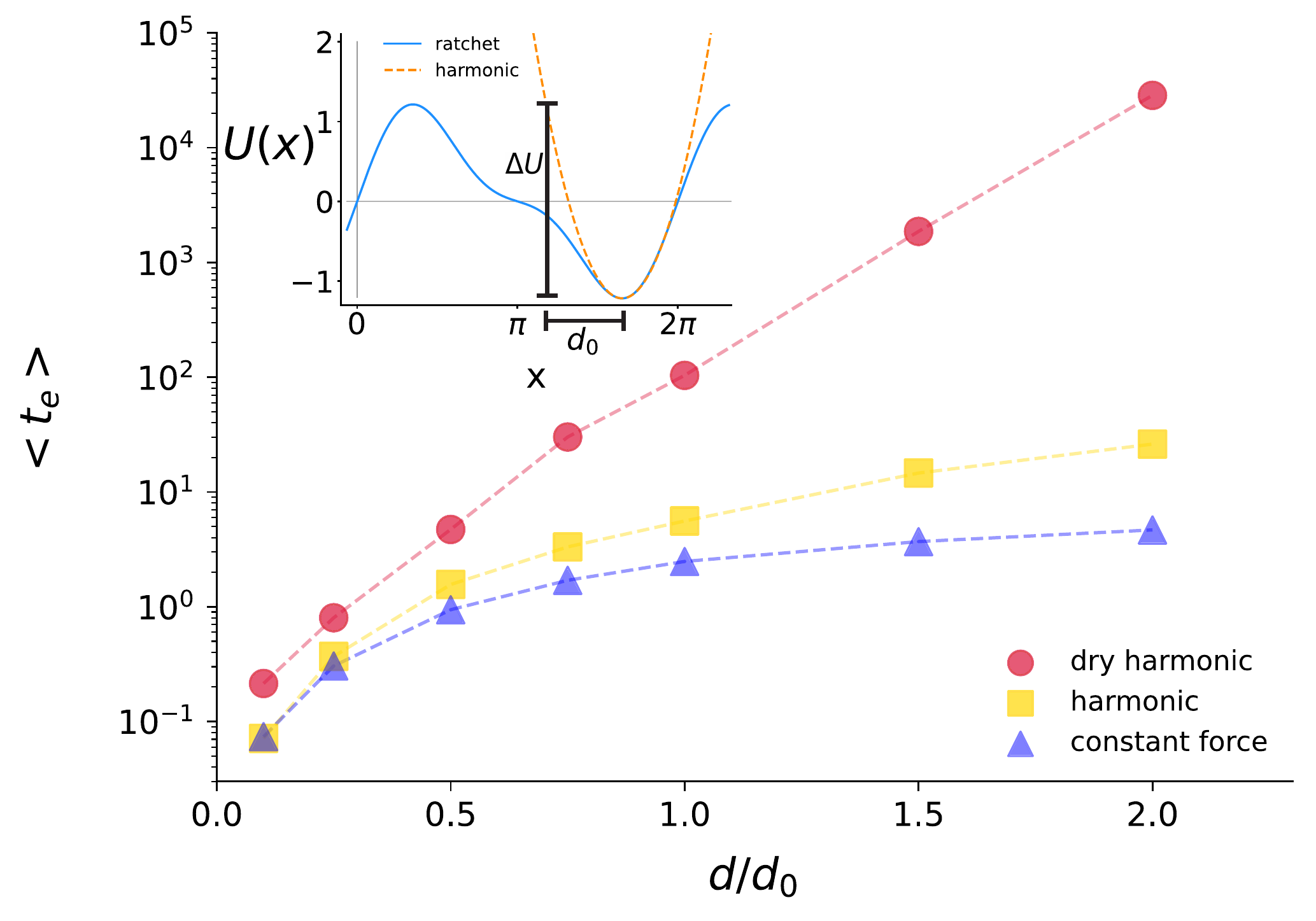}
\vspace{-0.25cm}
\caption{
\footnotesize{{\bf Average escape times.} Average escape time $\braket{t_e}$ for a harmonically confined Brownian particle  with and without Coulomb friction and constant-force starting at $x_0=0$ as a function of the right escape point
	. The inset reports a graphical depiction of the harmonic approximation introduced in the main text. For the sake of clarity, the harmonic potential is horizontally and graphically shifted. Parameters: $\gamma=0.05$, $T=10.0$, ($k=2.20$) ($\alpha=1.0$). $d_0=1.49$ is chosen in such a way that $kd_0^2/2=\Delta U$, with $\Delta U$ as the ratchet potential depth for $\mu=0.4$. $F=2.28\cdot 10^{-4}$ is the constant force corresponding to the $\gamma=0.05,T=10,\mu=0.4$ ratchet case.}}
\label{fig:fig5}
\end{figure}

Since the onset of the ratchet effect is related to the presence of the spatial potential, also affecting  the diffusion properties, it is interesting to further investigate the role of Coulomb friction on the escape time from a parabolic well. In particular, we compare the behaviour obtained in the ratchet model with that computed for the simple constant-force model. For the simple Langevin equation with no external force, analytical results are reviewed in~\cite{grebenkov2014first}. In the presence of dry friction and the absence of spatial potential, the problem has been addressed in~\cite{chen2014first}.
Here, {a harmonic approximation $kx^2/2$ of the ratchet potential around one if its minima is considered}. The elastic constant $k$ is obtained from a second-order expansion of the ratchet potential \eqref{potential} around one of its minima.
{The average time $\langle t_e\rangle$ necessary for the particle to reach a distance $d$ from the minimum of the potential was computed}. Distances are expressed in units of $d_0$, which was chosen in such a way that $kd_0^2/2=\Delta U$, with $\Delta U$ the ratchet potential depth (see the inset of Figure~\ref{fig:fig5} for a graphical depiction). We are interested in showing how the mean escape time $\braket{t_e}$ as function of the escape distance $d$ varies in several conditions. As shown in Figure~\ref{fig:fig5}, the smallest mean exit times are observed in the case of the constant force model. {More specifically, at small values of $d/d_0$, the behaviour for the constant force model is very similar to the harmonic case; both trends show a saturation, and significant differences only arise upon increasing the distance from the bottom of the well.  On the other hand, the effect of dry friction is much stronger, and the marked increase in exit times, in the range of explored parameters, seems to exhibit an exponential behaviour as a function of the distance $d/d_0$}. Finally, note that the very long exit times in the model with dry friction can be related to the extended plateaus observed in the variance, where the particle remains trapped in the well for long times. 

\section{Conclusions}

\label{concl}

 {In this work, a stochastic differential equation featuring non-linear friction and an asymmetric spatial potential has been studied}. The system is out of equilibrium and shows the occurrence of the ratchet effect, namely a net average drift, with a non-monotonic magnitude as a function of the bath parameters, temperature, and viscous friction. {The diffusion properties of the model have also been investigated}, with a particular focus on the position variance and on the MSD, finding the occurrence of different regimes and different characteristic times separating such time regimes. Finally, {the analysis} proved that the diffusion properties of the ratchet model under scrutiny present strong  differences with respect to a simple Brownian particle with inertia under the action of an external constant force. Our study contributes by shedding light on some important dynamical features of systems characterized by the presence of both non-linear friction and fluctuations, which play a central role in many natural phenomena, { from biological molecular motors at the cellular scale to earthquakes and avalanches at geophysical scales, and in experimental applications, in particular for nano- and micro-friction such as for nanometer contacts in the context of micro- and nanomachines~\cite{vanossi2013colloquium}.} 
 
We plan to extend the study of this model to the framework of stochastic thermodynamics, addressing the interesting issues related to the definition of entropy production and fluctuating efficiency, in future works.


\begin{thebibliography}{999}

\bibitem[H{\"a}nggi and Marchesoni(2009)]{hanggi2009artificial}
H{\"a}nggi, P.; Marchesoni, F.
\newblock Artificial Brownian motors: Controlling transport on the nanoscale.
\newblock {\em Rev. Mod. Phys.} {\bf 2009}, {\em 81},~387.

\bibitem[Reimann and H{\"a}nggi(2002)]{reimann2002introduction}
Reimann, P.; H{\"a}nggi, P.
\newblock Introduction to the physics of Brownian motors.
\newblock {\em Appl. Phys. A} {\bf 2002}, {\em 75},~169--178.

\bibitem[Dialynas \em{et~al.}(1997)Dialynas, Lindenberg, and
  Tsironis]{dialynas1997ratchet}
Dialynas, T.; Lindenberg, K.; Tsironis, G.
\newblock Ratchet motion induced by deterministic and correlated stochastic
  forces.
\newblock {\em Phys. Rev. E} {\bf 1997}, {\em 56},~3976.

\bibitem[Gradenigo \em{et~al.}(2010)Gradenigo, Sarracino, Villamaina, Grigera,
  and Puglisi]{gradenigo2010ratchet}
Gradenigo, G.; Sarracino, A.; Villamaina, D.; Grigera, T.S.; Puglisi, A.
\newblock The ratchet effect in an ageing glass.
\newblock {\em J. Stat. Mech. Theory Exp.} {\bf
  2010}, {\em 2010},~L12002.

\bibitem[Cleuren and Eichhorn(2008)]{cleuren2008dynamical}
Cleuren, B.; Eichhorn, R.
\newblock Dynamical properties of granular rotors.
\newblock {\em J. Stat. Mech. Theory Exp.} {\bf
  2008}, {\em 2008},~P10011.

\bibitem[Costantini \em{et~al.}(2007)Costantini, Marconi, and
  Puglisi]{costantini2007granular}
Costantini, G.; Marconi, U.M.B.; Puglisi, A.
\newblock Granular Brownian ratchet model.
\newblock {\em Phys. Rev. E} {\bf 2007}, {\em 75},~061124.

\bibitem[Sarracino(2013)]{sarracino2013time}
Sarracino, A.
\newblock Time asymmetry of the Kramers equation with nonlinear friction:
  Fluctuation-dissipation relation and ratchet effect.
\newblock {\em Phys. Rev. E} {\bf 2013}, {\em 88},~052124.

\bibitem[R and Benjamin(2022)]{arjun2022}
Benjamin, R.
\newblock Current and diffusion of Overdamped Active Brownian Particles in a
  Ratchet Potential. \emph{arXiv} \textbf{2022}, arXiv:2211.04298.


\bibitem[De~Gennes(2005)]{de2005brownian}
De~Gennes, P.G.
\newblock Brownian motion with dry friction.
\newblock {\em J. Stat. Phys.} {\bf 2005}, {\em 119},~953--962.

\bibitem[Hayakawa(2005)]{hayakawa2005langevin}
Hayakawa, H.
\newblock Langevin equation with Coulomb friction.
\newblock {\em Phys. D Nonlinear Phenom.} {\bf 2005}, {\em 205},~48--56.

\bibitem[Baule \em{et~al.}(2009)Baule, Cohen, and Touchette]{baule2009path}
Baule, A.; Cohen, E.; Touchette, H.
\newblock A path integral approach to random motion with nonlinear friction.
\newblock {\em J. Phys. Math. Theor.} {\bf 2009},
  {\em 43},~025003.

\bibitem[Touchette \em{et~al.}(2010)Touchette, Van~der Straeten, and
  Just]{touchette2010brownian}
Touchette, H.; Van~der Straeten, E.; Just, W.
\newblock Brownian motion with dry friction: Fokker--Planck approach.
\newblock {\em J. Phys. Math. Theor.} {\bf 2010},
  {\em 43},~445002.

\bibitem[Baule \em{et~al.}(2010)Baule, Touchette, and Cohen]{baule2010stick}
Baule, A.; Touchette, H.; Cohen, E.
\newblock Stick--slip motion of solids with dry friction subject to random
  vibrations and an external field.
\newblock {\em Nonlinearity} {\bf 2010}, {\em 24},~351.

\bibitem[Pototsky and Marchesoni(2013)]{pototsky2013periodically}
Pototsky, A.; Marchesoni, F.
\newblock Periodically driven Brownian motion with dry friction and
  ultrarelativistic Langevin equations.
\newblock {\em Phys. Rev. E} {\bf 2013}, {\em 87},~032132.

\bibitem[Feghhi \em{et~al.}(2022)Feghhi, Tichy, and Lau]{feghhi2022pulling}
Feghhi, T.; Tichy, W.; Lau, A.
\newblock Pulling a harmonically bound particle subjected to Coulombic
  friction: A nonequilibrium analysis.
\newblock {\em Phys. Rev. E} {\bf 2022}, {\em 106},~024407.

\bibitem[Cerino and Puglisi(2015)]{cerino2015entropy}
Cerino, L.; Puglisi, A.
\newblock Entropy production for velocity-dependent macroscopic forces: The
  problem of dissipation without fluctuations.
\newblock {\em EPL (Europhys. Lett.)} {\bf 2015}, {\em 111},~40012.

\bibitem[Sarracino \em{et~al.}(2013)Sarracino, Gnoli, and
  Puglisi]{sarracino2013ratchet}
Sarracino, A.; Gnoli, A.; Puglisi, A.
\newblock Ratchet effect driven by Coulomb friction: the asymmetric Rayleigh
  piston.
\newblock {\em Phys. Rev. E} {\bf 2013}, {\em 87},~040101.

\bibitem[Manacorda \em{et~al.}(2014)Manacorda, Puglisi, and
  Sarracino]{manacorda2014coulomb}
Manacorda, A.; Puglisi, A.; Sarracino, A.
\newblock Coulomb friction driving Brownian motors.
\newblock {\em Commun. Theor. Phys.} {\bf 2014}, {\em
  62},~505.

\bibitem[Gnoli \em{et~al.}(2013{\natexlab{a}})Gnoli, Petri, Dalton, Pontuale,
  Gradenigo, Sarracino, and Puglisi]{gnoli2013brownian}
Gnoli, A.; Petri, A.; Dalton, F.; Pontuale, G.; Gradenigo, G.; Sarracino, A.;
  Puglisi, A.
\newblock Brownian ratchet in a thermal bath driven by coulomb friction.
\newblock {\em Phys. Rev. Lett.} {\bf 2013}, {\em 110},~120601.

\bibitem[Gnoli \em{et~al.}(2013{\natexlab{b}})Gnoli, Puglisi, and
  Touchette]{gnoli2013granular}
Gnoli, A.; Puglisi, A.; Touchette, H.
\newblock Granular Brownian motion with dry friction.
\newblock {\em EPL (Europhys. Lett.)} {\bf 2013}, {\em 102},~14002.

\bibitem[Dubkov \em{et~al.}(2009)Dubkov, H{\"a}nggi, and
  Goychuk]{dubkov2009non}
Dubkov, A.; H{\"a}nggi, P.; Goychuk, I.
\newblock Non-linear Brownian motion: the problem of obtaining the thermal
  Langevin equation for a non-Gaussian bath.
\newblock {\em J. Stat. Mech. Theory Exp.} {\bf
  2009}, {\em 2009},~P01034.

\bibitem[Platen and Bruti-Liberati(2010)]{platen2010}
Platen, E.; Bruti-Liberati, N.
\newblock {\em Numerical solution of stochastic differential equations with
  jumps in finance}; Springer Science \& Business Media:  Berlin/Heidelberg, Germany, 
  2010; Volume~64.

\bibitem[Lindenberg \em{et~al.}(2007)Lindenberg, Sancho, Lacasta, and
  Sokolov]{lindenberg2007dispersionless}
Lindenberg, K.; Sancho, J.M.; Lacasta, A.; Sokolov, I.M.
\newblock Dispersionless transport in a washboard potential.
\newblock {\em Phys. Rev. Lett.} {\bf 2007}, {\em 98},~020602.

\bibitem[Kaffashnia and Evstigneev(2021)]{kaffashnia2021origin}
Kaffashnia, A.; Evstigneev, M.
\newblock Origin of dispersionless transport in spite of thermal noise.
\newblock {\em Phys. Rev. E} {\bf 2021}, {\em 104},~054113.

\bibitem[Marchenko \em{et~al.}(2022)Marchenko, Aksenova, Marchenko, and
  Zhiglo]{marchenko2022dispersionless}
Marchenko, I.; Aksenova, V.Y.; Marchenko, I.; Zhiglo, A.
\newblock Dispersionless transport in washboard potentials revisited.
\newblock {\em J. Phys. Math. Theor.} {\bf 2022},
  {\em 55},~155005.

\bibitem[Grebenkov(2014)]{grebenkov2014first}
Grebenkov, D.S.
\newblock First exit times of harmonically trapped particles: a didactic
  review.
\newblock {\em J. Phys. A Math. Theor.} {\bf 2014},
  {\em 48},~013001.

\bibitem[Chen and Just(2014)]{chen2014first}
Chen, Y.; Just, W.
\newblock First-passage time of Brownian motion with dry friction.
\newblock {\em Phys. Rev. E} {\bf 2014}, {\em 89},~022103.

\bibitem[Vanossi \em{et~al.}(2013)Vanossi, Manini, Urbakh, Zapperi, and
  Tosatti]{vanossi2013colloquium}
Vanossi, A.; Manini, N.; Urbakh, M.; Zapperi, S.; Tosatti, E.
\newblock Colloquium: Modeling friction: From nanoscale to mesoscale.
\newblock {\em Rev. Mod. Phys.} {\bf 2013}, {\em 85},~529.

\end{thebibliography}
\end{document}